\documentclass[a4paper,11pt]{article}
\usepackage{pos}

\usepackage{slashed}

\renewcommand{\P}{{\cal P}}

\newcommand{\Q}{{\mathscr Q}}
\newcommand{\E}{{\mathscr E}}

\def\eq#1{{Eq.~(\ref{#1})}}

\def\fig#1{{Fig.~\ref{#1}}}

\def\sect#1{{Sect.~\ref{#1}}}

\def\vev#1{\left\langle #1\right\rangle}
\def\abs#1{\left| #1\right|}

\def\Im{\mbox{Im}\,}

\def\Tr{\mbox{Tr}\,}

\def\det{\mbox{det}\,}

\def\diag{\mbox{diag}\,}

\renewcommand{\bar}{\overline}

\newcommand{\beq}{\begin{equation}}
\newcommand{\eeq}{\end{equation}}
\newcommand{\bea}{\begin{eqnarray}}
\newcommand{\eea}{\end{eqnarray}}

\renewcommand{\[}{\left[}
\renewcommand{\]}{\right]}
\renewcommand{\(}{\left(}
\renewcommand{\)}{\right)}

\title{CP-violating Axions}

\author*[a]{Luca Di Luzio}

\affiliation[a]{Dipartimento di Fisica e Astronomia `G.~Galilei', Universit\`a di Padova, Italy \\
        Istituto Nazionale Fisica Nucleare, Sezione di Padova, Italy}

\emailAdd{luca.diluzio@unipd.it}

\abstract{While the axion was originally introduced to ``wash out'' CP violation from strong interactions, new sources of CP violation beyond QCD 
might manifest themselves via a tiny scalar axion-nucleon component. The latter can be experimentally probed in axion-mediated force experiments, as suggested long ago by J.E.~Moody and F.~Wilczek. In the present note, I review the 
physical origin of CP-violating axion couplings 
and 
point out the special role 
of the QCD axion as a low-energy portal  to high-energy sources of CP violation.
}

\FullConference{%
  The European Physical Society Conference on High Energy Physics (EPS-HEP2021),\\
  26-30 July 2021\\
  Online conference, jointly organized by Universität Hamburg and the research center DESY
}

\tableofcontents

\begin{document}
\maketitle



\section{QCD axion $\&$ CP}

The QCD axion originally emerged from the need to 
``wash out'' CP violation from strong interactions \cite{Peccei:1977hh,Peccei:1977ur,Weinberg:1977ma,Wilczek:1977pj}. 
From an effective field theory perspective, the Peccei-Quinn (PQ) solution to the strong CP problem can be formulated as follows. 
The SM Lagrangian is augmented by a spin-0 field $a(x)$ endowed  with a pseudo-shift symmetry $a \to a + \alpha f_a$, 
that is broken only by the operator 
\beq 
\label{eq:aGGtilde}
\frac{a}{f_a} \frac{\alpha_s}{8\pi}  G \tilde G 
\, , 
\eeq
with $G\tilde G \equiv \frac{1}{2} \epsilon^{\mu\nu\rho\sigma} G^a_{\mu\nu} G^a_{\rho\sigma}$. 
After employing the pseudo-shift 
symmetry to reabsorb the QCD $\theta$ term by setting $\alpha = - \theta$, one is left 
in the shifted theory
with the  
operator in \eq{eq:aGGtilde}. Hence, the question of CP violation in strong interactions 
is traded for 
a dynamical question about the vacuum expectation value (VEV) of the axion field
\beq 
\theta_{\rm eff} \equiv \frac{\vev{a}}{f_a} \, , 
\eeq
resulting
into an effective $\theta$ parameter,  
with 
$\abs{\theta_{\rm eff}} \lesssim 10^{-10}$ 
from the non-observation of the 
neutron electric dipol moment (nEDM). 

A general result, due to Vafa and Witten \cite{Vafa:1984xg} 
ensures that the ground state energy density 
of QCD $\E$ is minimized for $\theta_{\rm eff} = 0$,  
namely $\E(0) \leq \E (\theta_{\rm eff})$. 
The argument is based on an inequality that exploits the 
path-integral representation of $\E$ \cite{Coleman:1985rnk}. 
At large 4-volume in Euclidean space,\footnote{To see the origin of the 
$i$ factor in front of $\theta_{\rm eff}$, remember that 
$\epsilon^{\mu\nu\rho\sigma} G^a_{\mu\nu} G^a_{\rho\sigma} = 
\epsilon^{0123} G^a_{01} G^a_{23} + \ldots$ and the 
Wick rotation to Euclidean time $x_0 \to -i x_4$
implies $G^a_{0i} \to i G^a_{4i}$.} 
one has 
\beq
\label{eq:VF}
e^{-V_4 \E (\theta_{\rm eff})} = 
\int \mathcal{D} \varphi e^{-S_0 + i \theta_{\rm eff} \Q}  
= \Big| \int \mathcal{D} \varphi e^{-S_0 + i \theta_{\rm eff} \Q} \Big| 
\leq 
\int \mathcal{D} \varphi \Big| e^{-S_0 + i \theta_{\rm eff} \Q} \Big| 
= e^{-V_4 \E (0)} \, ,
\eeq
where $\varphi$ denotes the collection of QCD fields (quarks and gluons), 
$S_0$ is the QCD Lagrangian in absence of the $\theta$ term 
and we introduced the topological charge operator $\Q \equiv \int d^4x \frac{\alpha_s}{8\pi}  G \tilde G$, 
which takes integer values in the background of QCD instantons. 
A crucial assumption, on which the proof in \eq{eq:VF} relies on,
consists in the positive definiteness of the path-integral measure 
\beq 
\mathcal{D} \varphi \equiv \mathcal{D} A^a_\mu \, \det(\slashed{D} + M) \, , 
\eeq
where the gaussian path-integral over the fermionic fields has been explicitly performed. 
However, while the fermionic determinant is positive definite in vector-like theory like 
QCD,\footnote{Given the Euclidean Dirac operator $i\slashed{D} = i \sum_{\alpha=1}^4\gamma^\alpha D_\alpha$,  
the non-zero eigenvalues $\lambda$ of $i\slashed{D}$ are such that if 
$i\slashed{D} \psi = \lambda \psi$, then $i\slashed{D} (\gamma_5 \psi) = - \lambda (\gamma_5 \psi)$. 
So both $\lambda$ and $-\lambda$ are eigenvalues. 
Hence, being the Euclidean Dirac Lagrangian $\bar\psi (\slashed{D} + M) \psi$,
the fermionic determinant 
is 
$\det(\slashed{D} + M) = \Pi_{\lambda} (M - i \lambda)
= \Pi_{\lambda > 0} (M - i \lambda) (M + i \lambda) 
= \Pi_{\lambda > 0} (M^2 + \lambda^2) > 0$ 
(see e.g.~\cite{Vafa:1983tf}).} 
that is not the case 
for a chiral theory like the Standard Model (SM).   
Hence, we cannot apply the the Vafa-Witten theorem to the 
SM, although the argument does not automatically imply that $\theta_{\rm eff} \neq 0$ in the SM. 

An extra ingredient of the SM is that 
CP is explicitly broken in the quark sector by the Cabibbo-Kobayashi-Maskawa (CKM) phase, 
which sources a $\theta_{\rm eff} \neq 0$ term. To show this   
on general grounds, 
consider a CP transformation on all SM fields $\varphi \to \varphi'$. 
In absence of ultraviolet (UV) sources of CP violation one has 
$S_0 (\varphi) = S_0 (\varphi')$ and 
\begin{align}
\label{eq:VF2}
e^{-V_4 \E (\theta_{\rm eff})} &= 
\int \mathcal{D} \varphi e^{-S_0 (\varphi) + i \theta_{\rm eff} \Q (\varphi)} = 
\int \mathcal{D} \varphi' e^{-S_0 (\varphi') + i \theta_{\rm eff} \Q (\varphi')} \nonumber \\
&= \int \mathcal{D} \varphi e^{-S_0 (\varphi) - i \theta_{\rm eff} \Q (\varphi)}  
= e^{-V_4 \E (-\theta_{\rm eff})} \, ,
\end{align} 
where we have performed the change of variables $\varphi \to \varphi'$ in the path-integral (second step)
and taken into account the CP properties of the topological charge: $\Q(\varphi) = - \Q(\varphi')$ (third step).  
Hence, \eq{eq:VF2} shows that $\E$ is an even function: $\E (\theta_{\rm eff}) = \E (-\theta_{\rm eff})$. 
However, in the presence of UV sources of CP violation (like the CKM phase) one has 
$S_0 (\varphi) \neq S_0 (\varphi')$, which implies $\E (\theta_{\rm eff}) \neq \E (-\theta_{\rm eff})$. 
Therefore, since the energy density picks up an odd component in $\theta_{\rm eff}$, 
the minimum of $\E$ gets displaced from $\theta_{\rm eff} = 0$. 

Given a short-distance CP-violating local operator $\mathscr{O}_{\rm CPV} (x)$, 
we can estimate the value of $\theta_{\rm eff}$  
by expanding the axion potential as 
\beq 
\label{eq:Vthetaeff}
V(\theta_{\rm eff}) = K' \theta_{\rm eff} + \frac{1}{2} K \theta_{\rm eff}^2 + \mathcal{O} (\theta^3_{\rm eff}) \, , 
\eeq 
and focus on the $\theta_{\rm eff} \ll 1 $ regime, since we know 
that $\abs{\theta_{\rm eff}} \lesssim 10^{-10}$. 
Here, $K$ is a 2-point function also known as 
topological susceptibility \cite{Shifman:1979if}
\beq
\label{eq:defK}
K = i \int d^4 x 
\langle 0 | T \frac{\alpha_s}{8\pi} G\tilde G (x) \frac{\alpha_s}{8\pi} G\tilde G (0) | 0 \rangle 
\, , 
\eeq
while $K'$ is a 1-point function given by \cite{Pospelov:1997uv}
\beq
\label{eq:defKp}
K' = i \int d^4 x 
\langle 0 | T \frac{\alpha_s}{8\pi} G\tilde G (x) \mathscr{O}_{{\rm CPV}} (0) | 0 \rangle 
\, . 
\eeq
Note that $K' \neq 0$ because $G\tilde G$ and $\mathscr{O}_{{\rm CPV}}$ are both CP-odd, 
and CP-violating effects can be safely neglected for the evaluation of the QCD matrix element. 
The induced $\theta_{\rm eff}$ is hence obtained by the direct minimization of $V(\theta_{\rm eff})$ in \eq{eq:Vthetaeff}, yielding 
\beq
\label{eq:thetaeffmin}
\theta_{\rm eff} \simeq - \frac{K'}{K} \, .
\eeq

\section{$\theta_{\rm eff}$ in the Standard Model}

The value of $\theta_{\rm eff}$ in the SM was estimated by Georgi and Randall in Ref.~\cite{Georgi:1986kr}. 
At energies below $\Lambda_{\chi} = 4\pi F_\pi$ with $F_\pi \simeq 92$ MeV one can write a flavour conserving, 
CP-violating operator
\beq 
\mathscr{O}^{\rm SM}_{\rm CPV} = \frac{G_F^2}{m_c^2} J_{\rm CKM} 
[\bar u \gamma^\mu (1-\gamma_5) d \cdot \bar d \gamma_\mu] \slashed{D} 
[\gamma^\nu (1-\gamma_5) s \cdot\bar s \gamma_\nu (1-\gamma_5) u] \, ,
\eeq
which is obtained in the SM after integrating out the $W$ boson and the charm quark 
(see diagram in Fig.~1 of \cite{Georgi:1986kr}),  
while $J_{\rm CKM} = \Im V_{ud}V^*_{cd}V_{cs}V^*_{us} \simeq 3 \times 10^{-5}$ is the reduced Jarlskog invariant \cite{Jarlskog:1985ht}. 
A proper evaluation of the $K'$ matrix element in the presence of $\mathscr{O}^{\rm SM}_{\rm CPV}$ is far from trivial. 
However, by the rules of naive dimensional analysis (NDA) \cite{Manohar:1983md}
one expects $K' \sim F_\pi^{4} \frac{G_F^2}{m_c^2} J_{\rm CKM} F_\pi^{4} \Lambda^2_\chi$ 
\cite{Georgi:1986kr}.\footnote{The extension of NDA power counting 
arguments to CP-violating operators beyond the SM can be found in Ref.~\cite{Barbieri:1996vt}.} 
Hence, taking 
also 
$K \sim F_\pi^4$ (in reality 
$K \simeq (76 \ \text{MeV})^4$ \cite{diCortona:2015ldu,Borsanyi:2016ksw}),  
one obtains the estimate 
\beq
\label{eq:thetaeffmin2}
\theta^{\rm SM}_{\rm eff} \sim \frac{G_F^2}{m_c^2} J_{\rm CKM} F_\pi^4 \Lambda^{2}_{\chi} \sim 10^{-19} \, ,
\eeq
which should be taken only as indicative, since it
could be off also at the order of magnitude level. 
Anyway, 
the estimate in \eq{eq:thetaeffmin2} 
leads to two 
observations: 
\begin{enumerate}
\item \emph{The PQ mechanism works in the SM because the SM is the SM.} 

Indeed, the CKM contribution to 
$\theta^{\rm SM}_{\rm eff}$ could have easily overshoot $10^{-10}$ if the QCD and the Fermi scale would have been 
closer 
and/or in the presence of a trivial flavour structure such that $J_{\rm CKM} \sim 1$. 
Also, the PQ mechanism is not generically going to work in a low-scale theory beyond the SM 
with generic CP violation. To see this, assume a $d=6$ CP-violating operator 
$\mathscr{O}_{\rm CPV}/\Lambda_{\rm BSM}^2$, 
coupled to QCD.  
Following an estimate similar to that in \eq{eq:thetaeffmin2} one obtains 
\beq 
\label{eq:thetaeffmin3}
\theta^{\rm BSM}_{\rm eff} \sim \( \frac{\Lambda_\chi}{\Lambda_{\rm BSM}} \)^2 \sim 10^{-10} 
\( \frac{100 \ \text{TeV}}{\Lambda_{\rm BSM}} \)^2 \, , 
\eeq
which shows that the axion is not going to solve the strong CP problem\footnote{More precisely, 
the nEDM is going to set a constraint on $\Lambda_{\rm BSM}$ through the axion VEV.} 
in the presence of generic CP-violating new physics at the scale $\Lambda_{\rm BSM} \lesssim 100$ TeV. 

\item \emph{A no-lose theorem for the SM + axion?}

From a static measurement 
of the nEDM, it would be in principle possible to disentangle 
the axion contribution (using \eq{eq:thetaeffmin2})
\beq
d_n^{\rm SM \, + \, axion} \simeq 10^{-16} \theta^{\rm SM}_{\rm eff} \, e \, \text{cm} \sim 10^{-35}  \, e \, \text{cm} \, , 
\eeq
from the CKM-induced SM one (see \cite{Pospelov:2005pr} and references therein)
\beq
d_n^{\rm SM} \sim 10^{-32} \, e \, \text{cm} \, ,  
\eeq
which should be compared with the current experimental sensitivity $|d_n^{\rm exp}| \lesssim 10^{-26} \, e \, \text{cm}$. 
Although this strategy sounds extremely unlikely  
due to the huge improvements that are needed both from the experimental and the theoretical point of view,  
it is a conceptually interesting one, since it would be a direct test of the axion ground state.    
This also leads to the next relevant question: is there another way to test the axion ground state $\theta_{\rm eff}$?

\end{enumerate}

\section{CP-violating axion-nucleon couplings}

A remarkable consequence of $\theta_{\rm eff} \neq 0$ is the generation of a \emph{scalar} 
axion coupling to nucleons, $g^S_{aN} a \bar N N$ (with $N= p, n$), 
which can be searched for in axion-mediated force experiments 
as suggested by Moody and Wilczek \cite{Moody:1984ba}.  
The phenomenological relevance of the latter will be discussed in \sect{sec:axionforces}. 
To understand the origin of $g^S_{aN}$, let us define the canonical axion field as the 
excitation over its VEV 
$a \to \theta_{\rm eff} f_a + a$ and consider the two-flavour QCD-axion Lagrangian 
\beq 
\label{eq:LaQCD}
\mathcal{L}_a = \frac{\alpha_s}{8\pi} \( \theta_{\rm eff} + \frac{a}{f_a} \) G \tilde G - \bar q_L M_q q_R + \text{h.c.} \, , 
\eeq
where $M_q = \diag (m_u, m_d)$. 
Upon the axial quark rotation 
\beq 
\label{eq:qaxial}
q \to e^{\frac{i}{2}\gamma_5 \( \theta_{\rm eff} + \frac{a}{f_a} \) Q_a} q \, ,
\eeq
with $Q_a = M_q^{-1} / \Tr M_q^{-1} = \diag (\frac{m_d}{m_u+m_d}, \frac{m_u}{m_u+m_d}) $, 
the $G\tilde G$ term gets rotated away in \eq{eq:LaQCD}, while the 
quark mass matrix becomes 
\beq 
\label{eq:Ma}
M_q \to M_a = e^{i\gamma_5 \( \theta_{\rm eff} + \frac{a}{f_a} \) Q_a} M_q \, , 
\eeq
where we used $[M_q, Q_a] = 0$. Hence, in the new basis we have 
\begin{align}
\label{eq:LaQCD2}
\mathcal{L}_a &= - \bar q \[ \frac{M_a + M^\dag_a}{2} 
+ \gamma_5 \frac{M_a - M^\dag_a}{2} \] q \nonumber \\
&= - \bar q  \cos \( \( \theta_{\rm eff} + \frac{a}{f_a} \) Q_a \) M_q q
- \bar q  i \gamma_5 \sin \( \( \theta_{\rm eff} + \frac{a}{f_a} \) Q_a \) M_q q
\, . 
\end{align}
Focussing on the scalar component (i.e.~the one without $\gamma_5$) 
\begin{align}
\label{eq:LaQCD2}
\mathcal{L}_a &\supset 
- \bar q \cos \( \theta_{\rm eff} Q_a \) \cos \( \frac{a}{f_a} Q_a \)  M_q q
+ \bar q \sin \( \theta_{\rm eff} Q_a \) \sin \( \frac{a}{f_a} Q_a \) M_q q \nonumber \\ 
&\simeq \theta_{\rm eff}  \frac{a}{f_a} \bar q \, Q_a^2  M_q q = 
\frac{\theta_{\rm eff}}{f_a} \frac{m_u m_d}{m_u + m_d} a 
\( \frac{m_d}{m_u + m_d} \bar u u + \frac{m_u}{m_u + m_d} \bar d d \) 
\, ,  
\end{align}
where in the second step we kept only a term linear in $a/f_a$ and expanded for small $\theta_{\rm eff}$, 
while in the last step we employed the explicit representation of $Q_a$ below \eq{eq:qaxial}. 
Hence, the scalar axion coupling to nucleons is 
\begin{align} 
g^S_{aN} &= \frac{\theta_{\rm eff}}{f_a} \frac{m_u m_d}{m_u + m_d} 
\( \frac{m_d}{m_u + m_d} \langle N | \bar u u | N \rangle + \frac{m_u}{m_u + m_d} \langle N | \bar d d | N \rangle \)  \\
&= \frac{\theta_{\rm eff}}{f_a} \frac{m_u m_d}{m_u + m_d}
\[ 
\frac{\langle N | \bar u u + \bar d d | N \rangle}{2}
+ \( \frac{m_d - m_u}{m_u + m_d} \) \frac{\langle N | \bar u u - \bar d d | N \rangle}{2} \] \nonumber \, ,  
\end{align}
where in the last step we isolated the iso-spin singlet component, 
$\bar u u + \bar d d$. 
Focussing on the leading iso-spin singlet term (as in \cite{Moody:1984ba}), one has  
\beq 
\label{eq:fromthetaefftogaN}
g^S_{aN} \simeq \frac{\theta_{\rm eff}}{f_a} \frac{m_u m_d}{m_u + m_d} 
\frac{\langle N | \bar u u + \bar d d | N \rangle}{2} 
\simeq 1.3 \cdot 10^{-12} \, \theta_{\rm eff} \( \frac{10^{10} \ \text{GeV}}{f_a} \) 
\, . 
\eeq 
For the numerical evaluation we employed the 
value 
of the 
pion-nucleon sigma term 
$\sigma_{\pi N} = \langle N | \bar u u + \bar d d | N \rangle (m_u + m_d) / 2 = 59.1 \pm 3.5$ MeV \cite{Hoferichter:2015dsa}. 
As pointed out in Ref.~\cite{Bertolini:2020hjc},  
a factor $1/2$ was missed 
in the original derivation of Moody and Wilczek \cite{Moody:1984ba}. 

\section{A new master formula for $g^S_{aN}$}
\label{sec:masterformula}

Although \eq{eq:fromthetaefftogaN} is sufficient for the sake of an estimate, it turns out to be conceptually 
unsatisfactory when we focus on the relevant question: \emph{how to properly 
impose the nEDM bound?} 
The reason being the it misses extra contributions due to meson tadpoles ($\pi^0$, $\eta$, $\eta'$), 
which are 
generated by the same UV sources of CP violation responsible for $\theta_{\rm eff} \neq 0$
and 
are of the same size of the one proportional to $\theta_{\rm eff}$. 
This improvement was recently taken into account in Ref.~\cite{Bertolini:2020hjc} which,  
including as well iso-spin breaking effects from the leading order (LO) axion-baryon-meson chiral Lagrangian, 
found\footnote{The importance of meson tadpole contributions was previously pointed out in \cite{Pospelov:1997uv}, 
while iso-spin breaking effects were also taken into account in \cite{Bigazzi:2019hav}.} 
\begin{align}
\label{gan-pVEV}
{g}^S_{an,\,p} &\simeq 
\frac{4B_0\, m_u m_d}{f_a (m_u+m_d)}  \bigg[\pm (b_D+b_F)\frac{\vev{\pi^0}}{F_\pi}
+ \frac{b_D-3b_F}{\sqrt{3}}\frac{\vev{\eta_8}}{F_\pi}  
 -\sqrt{\frac{2}{3}}(3b_0+2b_D) \frac{\vev{\eta_0}}{F_\pi} \nonumber \\
&-   \left(b_0 + (b_D+b_F)\frac{m_{u,d}}{m_d+m_u}\right)\theta_{\rm eff}\bigg] \, , 
\end{align}
where for clarity we neglected $m_{u,d}/m_s$ terms. Here, $B_0=m_\pi^2/(m_d+m_u)$ while the hadronic
Lagrangian parameters $b_{D,F}$ are determined from the baryon octet mass splittings,
$b_D\simeq 0.07\,\rm GeV^{-1}$, $b_F\simeq -0.21\,\rm GeV^{-1}$ at LO~\cite{Pich:1991fq}.  The value
of $b_0$ is determined from the pion-nucleon sigma term as $b_0\simeq -\sigma_{\pi N}/4m_\pi^2$.
From the determination in~\cite{Hoferichter:2015dsa} one obtains
$b_0\simeq -0.76\pm 0.04\, \rm GeV^{-1}$. 
Given $\sigma_{\pi N}\equiv \langle N | \bar u u + \bar d d | N \rangle\, (m_u+m_d)/2$, 
the isospin symmetric  $b_0\theta_{\rm eff}$ term reproduces exactly \eq{eq:fromthetaefftogaN}.

In general, $g^S_{aN}$ and $d_n$ are \emph{not} proportional, as it would follow instead 
 from~\eq{eq:fromthetaefftogaN}. For instance, exact cancellations among the VEVs can happen
for $d_n$~\cite{Cirigliano:2016yhc,Bertolini:2019out} which have no counterpart in $g^S_{aN}$.
Note that $\theta_{\rm eff}$ and the meson VEVs 
in \eq{gan-pVEV}  
are meant to be computed from a high-energy source of
CP violation, represented by an effective operator $\mathscr{O}_{\rm CPV}$. 
In Ref.~\cite{Bertolini:2020hjc} an explicit example 
was worked out in the context of 4-quark operators of the type 
$\mathscr{O}_{\rm CPV} = (\bar q q) (\bar q' i \gamma_5 q')$ 
with $q,q'= u,d,s$, 
arising e.g.~by integrating out the heavy $W_R$ boson in 
left-right symmetric models. Building 
on the detailed analysis of Ref.~\cite{Bertolini:2019out},  
both $g^S_{aN}$ and $d_n$ were computed in the minimal left-right symmetric model 
with $\P$-parity \cite{Senjanovic:1978ev,Mohapatra:1979ia}, 
showing a non-trivial interplay which deviates sizeably from the naive approach of 
\eq{eq:fromthetaefftogaN} with $g^S_{aN} \propto d_n \propto \theta_{\rm eff}$.


\section{Axion-mediated forces}
\label{sec:axionforces}

Including both scalar and pseudo-scalar couplings to matter fields,  
the axion interaction Lagrangian  can be written as\footnote{More general 
CP-violating axion-like particle interactions have been  
recently analyzed in \cite{DiLuzio:2020oah}.}  
$\mathcal{L}^{\rm int}_a = 
g^S_{aN} a \bar N N + 
g^P_{af} a \bar f i \gamma_5 f$ (here $f = N, e$),
where $g^P_{af} = C_f m_f / f_a$ is the usual pseudo-scalar axion coupling, 
with $C_f  \sim \mathcal{O}(1)$ in benchmark axion models \cite{DiLuzio:2020wdo}. 
By taking the non-relativistic limit of $\mathcal{L}^{\rm int}_a$ one obtains different kinds of static potentials,  
which can manifest themselves as new axion-mediated macroscopic forces \cite{Moody:1984ba}. 
The latter can be tested in laboratory experiments, and hence do not rely on model-dependent axion production mechanisms, 
as in the case of dark matter axions (haloscopes) or to a less extent solar axions (helioscopes). 
An updated review of axion-mediated force experiments and relevant limits can be found in 
Ref.~\cite{OHare:2020wah} (see also \cite{Raffelt:2012sp,Irastorza:2018dyq,Sikivie:2020zpn}). 

Axion-induced potentials can be of three types,  
depending on the combination of couplings involved, 
i.e.~$g^S_{aN} g^S_{af}$ (monopole-monopole),
$g^S_{aN} g^P_{af}$ (monopole-dipole)
and $g^P_{af} g^P_{af}$ (dipole-dipole). 
The idea of searching for dipole-dipole axion interactions   
in atomic physics 
is as old as the axion itself \cite{Weinberg:1977ma}. 
However, dipole-dipole forces turn out to be 
spin suppressed in the non-relativistic limit 
and suffer from large backgrounds from ordinary magnetic forces. 
Searches based on monopole-monopole interactions, 
like tests of gravity on macroscopic scales, are in principle much more powerful.  
However, under the theoretical prejudice that we are after the QCD axion, 
the $\theta_{\rm eff}^2$ suppression (given the nEDM bound) is such that 
current experiments are still some orders of magnitude far from testing the QCD axion.  

\begin{figure}[t]
\centering
\includegraphics[width=15cm]{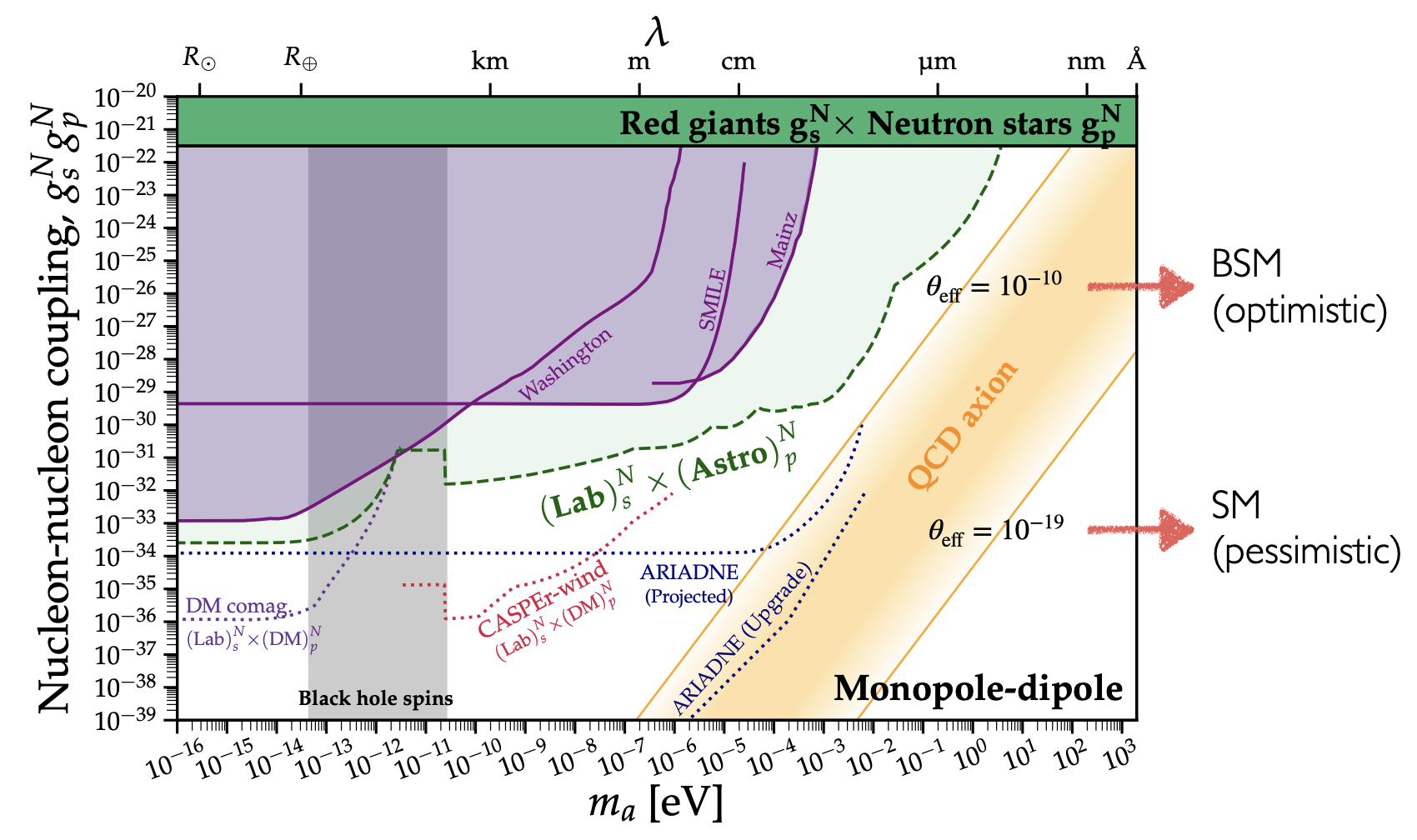}
\caption{Parameter space of axion-mediated monopole-dipole forces. Figure from Ref.~\cite{OHare:2020wah}.}
\label{fig:mondip}       
\end{figure}
The sweet spot is given by monopole-dipole searches which, as shown in \fig{fig:mondip}, 
will enter the QCD axion region in the near future. 
In fact, a new detection concept by the ARIADNE collaboration \cite{Arvanitaki:2014dfa,Geraci:2017bmq}
plans to use nuclear magnetic resonance techniques to probe the axion field sourced by an unpolarized material 
via a sample of nucleon spins.\footnote{A similar approach is pursued by the QUAX-$g_pg_s$ 
collaboration \cite{Crescini:2016lwj,Crescini:2017uxs} using instead electron spins.} 
Note that the yellow QCD axion band in \fig{fig:mondip} is obtained by 
employing \eq{eq:fromthetaefftogaN} for the scalar axion-nucleon coupling, 
with the value of $\theta_{\rm eff}$ spanning from the SM estimate 
$\theta_{\rm eff} \sim 10^{-19}$ 
to the limit imposed by the nEDM  
$\theta_{\rm eff} \simeq 10^{-10}$. 
However, as argued in \sect{sec:masterformula}, the relation between $g^S_{aN}$ and $d_n$ 
is model dependent and hence, given a specific source of CP-violation,  
it should be assessed case by case in order to properly determine the parameter space region 
that is allowed by the nEDM bound. 

\section{Outlook}

In order to have a testable signal in monopole-dipole axion searches 
a sizeable sources of CP violation beyond the SM is required. 
As a reference value in terms of $\theta_{\rm eff}$ 
(using \eq{eq:fromthetaefftogaN} for $g^S_{aN}$) 
one would need 
$\theta_{\rm eff} \gtrsim 10^{-13}$, that is 
three orders of magnitude below the 
current nEDM bound. 
Since CP is not a symmetry of nature, there is no 
reason to expect $\theta_{\rm eff} \to 0$.  
The SM itself predicts $\theta_{\rm eff} \sim 10^{-19}$, 
that is 
far from being testable. 
However,
new sources of CP violation 
beyond the CKM phase
are needed to explain the 
matter-antimatter asymmetry of the universe
and, even if 
decoupled 
at 
scales 
as heavy as 
100 TeV,  
they might contribute sizeably to $\theta_{\rm eff}$ 
(see estimate in \eq{eq:thetaeffmin3}).  

If an axion relaxation mechanism is at play,
then 
axion-mediated forces provide an alternative experimental handle 
for probing UV sources of CP violation in the quark sector, 
with projected sensitivities that are stronger than current EDM searches.\footnote{Future improvements 
on EDM limits 
might as well play a crucial role for disentangling 
different sources of CP violation coupled to QCD, 
also in the presence of a light axion field \cite{deVries:2021sxz}.} 
This suggests to rethink the role of the QCD axion 
from a ``laundry detergent'' of CP violation in the strong interactions 
to 
a low-energy portal to high-energy sources of CP violation, 
thus turning the 
strong CP problem into the strong CP opportunity.

\section*{Acknowledgments}

I thank
Stefano Bertolini,  
Giacomo Landini
and Fabrizio Nesti 
for pleasant discussions 
and collaboration on the topics 
of this note.




\bibliographystyle{JHEP}
\bibliography{bibliography}

\end{document}